\documentstyle[11pt,paspconf,epsf]{article}

\begin{document}
\newcommand{\hmpc}{h^{-1}\;{\rm Mpc}}
\newcommand{\lya}{Ly$\alpha$\ }
\newcommand{\overden}{{\rho/{\overline\rho}}}
\newcommand{\K}{{\rm K}}
\newcommand{\kms}{{\rm km}\;{\rm s}^{-1}}
\newcommand{\Nh}{N_{\rm HI}}
\newcommand{\nh}{n_{\rm HI}}
\newcommand{\cdunits}{{\rm cm}^{-2}}

\title{Simulating Cosmic Structure Formation}

\author{David H. Weinberg}
\affil{Department of Astronomy, Ohio State University, Columbus, OH 43210}

\author{Neal Katz}
\affil{Department of Physics and Astronomy, University of Massachusetts,
Amherst, MA 01003}

\author{Lars Hernquist}
\affil{Lick Observatory, University of California, Santa Cruz, CA 95064}

\begin{abstract}
We describe cosmological simulation techniques and their application
to studies of
cosmic structure formation, with particular attention to recent
hydrodynamic simulations of structure in the high redshift universe.
Collisionless N-body simulations with Gaussian initial conditions
produce a pattern of sheets, filaments,
tunnels, and voids that resembles the observed large scale galaxy
distribution.  Simulations that incorporate gas dynamics and dissipation
form dense clumps of cold gas with sizes and masses similar to the
luminous parts of galaxies.  Models based on inflation and cold dark
matter predict a healthy population of high redshift galaxies,
including systems with star formation rates of $20 M_\odot\;{\rm yr}^{-1}$
at $z=6$.  At $z \sim 3$, most of the baryons in these models
reside in the low density intergalactic medium, which produces fluctuating
\lya absorption in the spectra of background quasars.
The physical description of this ``\lya forest'' is particularly simple if 
the absorption spectrum is viewed as a 1-dimensional map of a continuous medium
instead of a collection of lines.  The combination of superb observational
data and robust numerical predictions makes the \lya forest a promising
tool for testing cosmological models.
\end{abstract}


\keywords{galaxies: formation, quasars: absorption lines, intergalactic
medium, large scale structure of Universe}

\section{Introduction}

The smoothness of the cosmic microwave background (CMB) tells us that the
early universe was remarkably homogeneous, a cosmos without galaxies, stars,
planets, or astronomers to admire them.  
Today's universe, on the other hand, exhibits 
structure over a vast range of scales.  How did this transition take place?
The leading hypothesis, strongly supported by COBE's discovery of 
anisotropies in the CMB (see Ned Wright's contribution to these proceedings),
is that gravitational instability amplified tiny fluctuations present in the
early universe into the rich structure that we observe today.
This broad hypothesis leaves many more specific questions
unanswered.  What were the properties of the primordial fluctuations,
and what was their physical origin?  What is the dark matter?
What are the values of the mass density parameter, $\Omega$,
and the cosmological constant (a.k.a.\ vacuum energy density),
$\Lambda$?

In principle, a set of answers to these questions, together with values
of parameters like $H_0$ and the baryon density parameter $\Omega_b$,
constitutes a theory of structure formation.  However, there is a large gap 
between a theory specified at this level and a set of observationally testable 
predictions.  Numerical simulations that start from the theoretically
specified initial conditions and evolve them forward in time
play an essential role in bridging this gap.
They show whether and how a theoretical model can produce objects like 
galaxies, quasars, and quasar absorption systems.
They often provide greater understanding
of observational phenomena, since in the simulations one knows 
the physical conditions, geometry, and history of the objects that form
and the relation between the observable tracers of structure and
the underlying distribution of dark matter.
This understanding is specific to a theoretical model or
class of theoretical models, but to the extent that the simulations
reproduce the observations, one may be tempted to believe at least
the general features of the physical picture that they provide,
even if it is not correct in all of its details.
By systematically exploring a range of models, one can see how changes
in initial conditions, cosmological parameters, or physical assumptions
reveal themselves in properties of observable structure.  

While cosmological simulations are often complex and computationally
intensive, some aspects of the results may be simple to understand,
and in such cases the simulations often suggest new analytic approximations,
whose accuracy and range of validity can be tested against the numerical 
results.  Simulations can also guide the development of new quantitative 
tools for characterizing observational data, providing physical motivation
for the methods and showing what properties of the underlying structure 
are constrained by different measures.  Finally, simulations yield
quantitative predictions for these statistical measures.  By
comparing them to observational data, one can test the general physical 
picture of structure formation that emerges from the simulations, and 
one can distinguish between cosmological models that
make different assumptions about primordial fluctuations, dark matter,
and the values of $\Omega$ and $\Lambda$.

In this paper we will describe several applications of cosmological
simulations, focusing on our recent work
using simulations with gas dynamics to investigate galaxy formation
and the \lya forest at $z \sim 2-6$.

\section{N-body simulations and large scale structure}

N-body simulations are the workhorses of numerical cosmology.
The basic technique is to represent the mass distribution in
some region of the universe with a
set of discrete particles, which start from an approximately
uniform distribution (usually a grid, occasionally a Poisson
distribution, or, more recently, a ``glass'' configuration
[\cite{white96}] in which the particle distribution is irregular
but the gravitational potential is uniform).
A simulation begins at high redshift, using the Zel'dovich (1970)
approximation to perturb the particles away from this uniform distribution
according to a random realization of the fluctuations predicted by a 
theoretical model.  The N-body code then integrates the equations
of motion forward in discrete timesteps, computing the gravitational potential
from the particle distribution, the particle accelerations
from the potential gradients, the updated particle velocities
from the accelerations, and the updated particle positions
from the velocities.  Gravitational forces are softened on
small scales in order to suppress two-body scattering, since the
particles are supposed to represent a collisionless fluid rather than
discrete objects.  The limits of a simulation's dynamic range
are set on the upper end by the box size (the typical simulation
volume is a periodic cube whose size remains fixed in {\it comoving}
coordinates) and on the lower end by the particle mass 
and the scale of the gravitational force softening.
Depending on the application,
the mass resolution or the force resolution 
may be the more important limiting factor.

Cosmological N-body algorithms differ mainly in the way that they
compute the gravitational accelerations from the particle distribution.
For sensible choices of numerical parameters, different codes generally
give consistent results on resolved scales when they are started
from the same initial conditions.  The gravitational clustering of matter 
in an expanding universe appears to be a ``forgiving'' numerical problem,
mainly because non-linear gravitational collapse effectively transfers 
information from large scales to small scales (\cite{lwp91}).  
As a result, numerical errors on small scales do not propagate 
upwards to affect large scale clustering; instead, these errors
are themselves ``overwritten'' when larger scale structures collapse.

Early work with N-body simulations showed that gravitational instability
models with plausible initial conditions could reproduce, at least
approximately, many of the basic observed properties of galaxy clustering,
such as the correlation function, the group multiplicity function,
and the large scale pattern of voids, sheets, and filaments
(e.g., \cite{gott79}; \cite{bhavsar81}; \cite{efstathiou81}; 
\cite{centrella83}; \cite{klypin83}).
By the mid-1980s, the focus of N-body work
had moved to quantitative measures of large
scale structure in different cosmological models, with the aim of
distinguishing 
between hot dark matter and cold dark matter and between an $\Omega=1$
universe and an open universe (e.g., \cite{white83}; \cite{davis85};
\cite{fry85}; \cite{melott87}).
Most of these studies adopted the general picture proffered by the
most natural versions of inflationary cosmology: adiabatic, Gaussian 
primordial fluctuations with a scale-invariant power spectrum.
However, there were also some investigations of models with 
other mechanisms for generating fluctuations, such as
cosmic strings (\cite{melott87a}; \cite{scherrer89}) or textures
(\cite{gooding92}).

As computers have become faster and codes more widely available,
N-body simulations have moved from a numerical specialty to a 
basic tool of the cosmological trade.
The ability to run large numbers of simulations at
reasonable computational expense has made it possible
to carry out systematic investigations 
of the parameter space of initial fluctuations and cosmological
parameters (e.g., \cite{efstathiou88}; \cite{wc92}; \cite{melott93})
and to create artificial redshift catalogs to aid the assessment
of systematic and statistical errors in analyses of observational
data ({\it numerous} examples, such as \cite{fisher94}).
Other recent work with N-body simulations includes studies of
the density profiles and survival of substructure in dark matter halos 
(e.g., \cite{navarro96}; \cite{moore96})
and tests of analytic descriptions of halo merger histories 
(e.g., \cite{lacey94}).

\begin{figure}
\centering
\centerline{\epsfxsize= 5.0 truein \epsfbox[20 0 545 750]{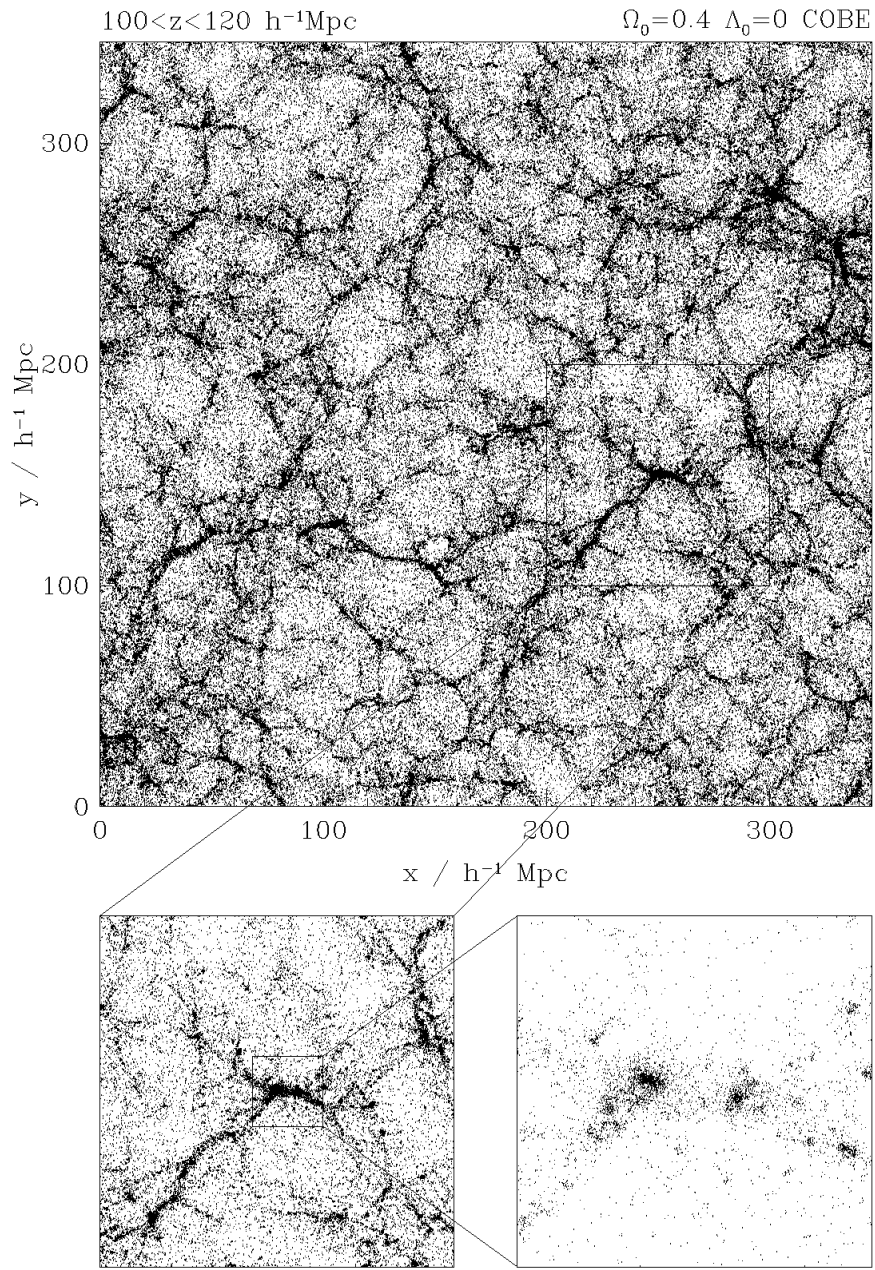}}
\caption{The particle distribution in a $20\hmpc$ thick slice
from an N-body simulation of an $\Omega_0=0.4$, open universe CDM model.
The large panel
shows the full cross section of the $360\hmpc$ simulation box, while the two
lower panels show successively expanded views, $100\hmpc$ and $20\hmpc$
on a side respectively. From Cole et al.\ (1997).  }
\label{figCdm}
\end{figure}

Figure~\ref{figCdm}, from Cole et al.\ (1997),
illustrates the kind of N-body simulation that
can now be run in a few days on a fast workstation.
The simulation represents a $360\hmpc$ cube of an open universe 
($\Omega_0=0.4$, $h \equiv H_0/100\;\kms\;{\rm Mpc}^{-1}=0.65$)
dominated by cold dark matter (CDM).
The initial fluctuations are Gaussian, as predicted by inflation,
and their amplitude is normalized to the amplitude of CMB
anisotropies observed by COBE.
Non-linear gravitational evolution transforms the Gaussian
initial fluctuations into a network of rounded voids and tunnels
interleaved with sheets and filaments.
The expanded views show that the extended high density structures
are themselves made up of smaller dark matter clumps.
Simulations that mimic the geometry and selection parameters
of galaxy redshift surveys show a good qualitative match between
the predictions of CDM models and the observed morphology of
large scale structure (e.g., \cite{white87}; \cite{park90}; \cite{park91}).
The qualitative and semi-quantitative demonstration that non-linear
gravitational evolution of ``generic'' Gaussian initial conditions
with a reasonable power spectrum can reproduce the basic observed
properties of large scale structure is probably the most important
general result to emerge from cosmological N-body studies.

The weakness of N-body simulations is that they do not show
where, when, or even whether galaxies form.  The smaller dark
matter clumps presumably correspond to the halos of individual
galaxies, but additional assumptions are required to assign luminosities 
to these halos.  More seriously, gravitational collapse and 
tidal disruption erase substructure in the cores of virialized objects.
By virtue of their masses, the largest halos in N-body simulations like 
the one shown in Figure~\ref{figCdm} should correspond to rich
galaxy clusters, but the inner regions of these halos are smooth,
without multiple condensations that could correspond to the halos of
individual galaxies.  The rapidity with which substructure is erased
by virialization depends to some extent on the resolution of the
N-body simulation (see \cite{gelb94}; \cite{moore96}), but
at bottom the problem appears to be not
numerical but physical: gravitational dynamics alone cannot
explain the existence of galaxy groups and clusters
(\cite{white78}; but see \cite{klypin97} for an alternative view).

\section{Gas dynamics, dissipation, and galaxy formation}

In order to model galaxy formation, cosmological simulations must include
additional physics that affects baryonic matter.
In keeping with currently popular theories of structure formation,
cosmological simulations with gas dynamics usually include a collisionless
dark matter component, treated by the N-body technique described in \S 2.
The dark matter interacts gravitationally with the gas component,
which also responds dynamically to shocks and pressure gradients.
The thermal state of the gas is influenced by adiabatic compression
and expansion, by entropy generation in shocks, by radiative cooling,
by photoionization heating from the background radiation field,
and, at high redshift ($z \ga 7$), by Compton cooling against the CMB.

Eulerian hydrodynamic codes solve the gas equations by finite difference methods
on a fixed mesh.  These types of codes have been widely used in other
areas of astronomy, physics, and engineering, and they are able to take
advantage of sophisticated schemes for capturing shocks.
Their principal drawback for studies of galaxy formation is the
fixed spatial mesh; with existing computers, it is impractical to
achieve resolution of a few kpc in a cosmologically interesting volume
(several Mpc or more on a side).
Lagrangian mesh codes or spatially adaptive Eulerian mesh codes can potentially
achieve high resolution in collapsed regions without requiring
enormous computational grids.
An alternative way to achieve spatially adaptive resolution is 
smoothed-particle hydrodynamics (SPH; \cite{gingold77}; \cite{lucy77}).
SPH codes represent the gas with particles; densities and pressure
gradients are computed by averaging (``smoothing'') over neighboring particles.
The smoothing is usually carried out over a fixed number of neighbors,
so the spatial resolution of the hydrodynamic calculations automatically
increases in the high density regions where it is most needed.
The strength of SPH for cosmological applications lies in this
high dynamic range.  The use of particles rather than a grid also
makes the technique flexible and ensures that dark matter and 
gas are modeled with the same gravitational resolution.
The code that we use for our simulations is TreeSPH 
(\cite{hernquist89}; \cite{kwh96}, hereafter KWH), 
which uses a hierarchical tree
code (\cite{barnes86}; \cite{hernquist87}) 
for gravitational force calculations and SPH for hydrodynamic calculations.

\begin{figure}
\centering
\centerline{\epsfxsize=5.1truein \epsfbox[70 180 565 760]{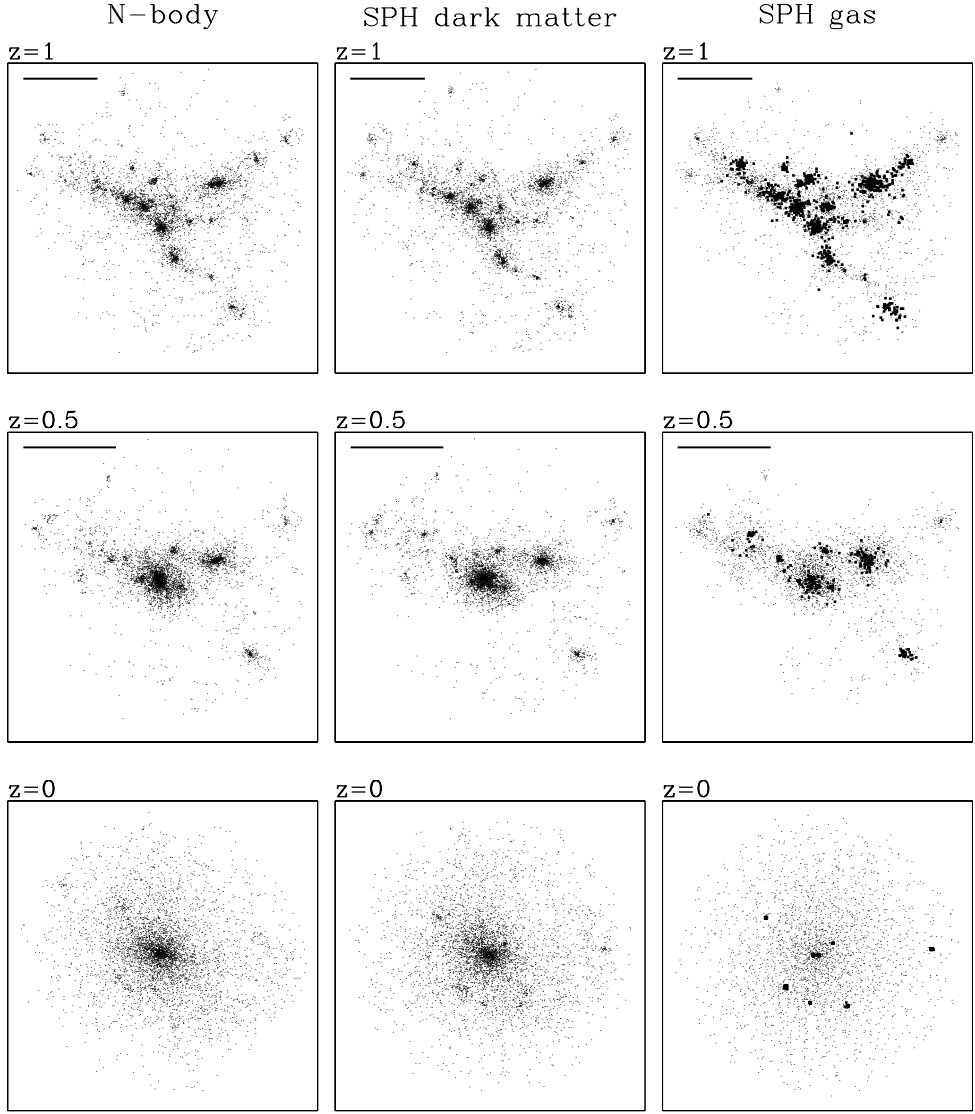}}
\caption{The formation of a small cluster of galaxies in 
N-body and SPH simulations of the SCDM model. 
The bottom panels show particles in a 1.25 Mpc sphere at $z=0$ in
the dissipationless N-body simulation (left), the dark component
of the SPH simulation (center), and the gas component of the SPH simulation
(right).  The cluster gas has a two-phase temperature-density structure;
heavy points show particles with $T < 30,000\;\K$, while
lighter points show hot particles, nearly all of which have $T > 10^6\;\K$.
The middle and upper rows show the same particles at $z=0.5$ and $z=1$,
with gas particles coded by their {\it final} ($z=0$) temperature.
The bottom panels are 2.7 Mpc on a side; horizontal lines in the upper 
panels indicate this 2.7 Mpc comoving scale at earlier redshifts.
}
\label{figCluster}
\end{figure}

Figure~\ref{figCluster} shows how dissipation in the gas component
works to resolve the ``overmerging'' problem raised at the end of \S 2.
The left hand panels are drawn from an N-body simulation 
(actually a TreeSPH simulation with no SPH component) of the
``standard'' CDM model (SCDM, with $\Omega=1$, $h=0.5$, and
an rms mass fluctuation in $8\hmpc$ spheres of $\sigma_8=0.7$).
The full simulation volume is a comoving cube 22 Mpc on a side,
but the figure shows only those particles that end up in a sphere
of radius 1.25 Mpc centered on the richest dark matter concentration.  
The mass in this sphere, $1.2 \times 10^{14} M_\odot$,
corresponds to the dark halo of a small cluster of galaxies.

At $z=1$, the particles that will form the cluster reside in a number of
smaller halos.  As gravitational clustering proceeds, these halos
merge, and the final product (lower left panel) is a single virialized
object, centrally concentrated but with little substructure.
Center panels show the dark matter component from an SPH simulation
started from the same initial conditions.
Right hand panels show the gas particles from the SPH simulation.
Heavier points have been used to represent gas particles that
are cold ($T < 30,000\;$K) at $z=0$; lighter points represent
hotter particles (most of which have 
$T \sim {\rm several} \times 10^6\;$K).

The evolution of the dark matter in the SPH simulation
is similar to that in the pure N-body simulation,
indicating that the gas component, with a mass fraction $\Omega_b=0.05$,
has little dynamical effect on the dark matter.
At $z=1$, the gas and dark matter distributions are also quite similar.
When the gas falls into the dark halos, it is shock heated to
roughly the virial temperature of the potential well, 
converting its kinetic infall energy into thermal energy.
However, some fraction of the gas is able to radiate its energy
by atomic cooling processes (see KWH for details); 
this gas loses pressure support and
settles into highly overdense knots at the halo centers.
These dense knots are much more resistant to disruption than the
relatively puffy halos that form in the dissipationless N-body simulation,
and they are able to survive as distinct entities even after their
parent halos merge to form the cluster.  At $z=0$, there are
eight of these dense, cold knots in the cluster, while the hot gas
is smoothly distributed with a density profile roughly like that 
of the dark matter.  The cold knots contain nearly 50\% of the
cluster's baryonic mass, though they look rather inconspicuous in
Figure~\ref{figCluster} because they are only $\sim 20$ kpc in size.
Inspection of the bottom panels shows that the baryons,
after following the dark matter's directions about where to collapse,
do have a small ``back reaction'': the gravity of the
condensed gas draws in the nearby dark matter, and the knots in
the outer part of the cluster have retained small dark matter halos
that are not present in the purely gravitational simulation.
Further aspects of this simulation are discussed in
Katz, Hernquist, \& Weinberg (1992) and KWH.

The process illustrated in Figure~\ref{figCluster} is essentially
the one envisioned by White \& Rees in 1978,
but modeling this process numerically required enormous advances
in computer hardware and in algorithms for treating gravity and gas dynamics.
It was not until the early 1990s that the formation of cold gas
concentrations with masses, sizes, and overdensities comparable
to the luminous regions of observed galaxies was achieved in
hydrodynamic cosmological simulations (e.g., \cite{khw92}; \cite{evrard94}).
Even simulations like these do not resolve the internal structure
of the ``galaxies'' that form in them. 
However, simulations that zoom in on individual objects can
achieve the higher resolution needed to examine some of the
details of galaxy formation (e.g., \cite{katz91}; \cite{katz92}; 
Steinmetz \& M\"uller 1994, 1995; \cite{vedel94}; 
\cite{navarro95}; \cite{navarro97}).

\section{High redshift galaxies}

We now shift focus to high redshift, the scene of much of the most
exciting observational and theoretical activity over the last few
years.  A particularly dramatic development has been the discovery
of a large population of star-forming galaxies at $z \sim 2-4$,
by the Lyman-break search technique (see Max Pettini's and
Piero Madau's contributions to these proceedings).

\begin{figure}
\centering
\centerline{\epsfxsize=5.4truein \epsfbox[105 235 535 700]{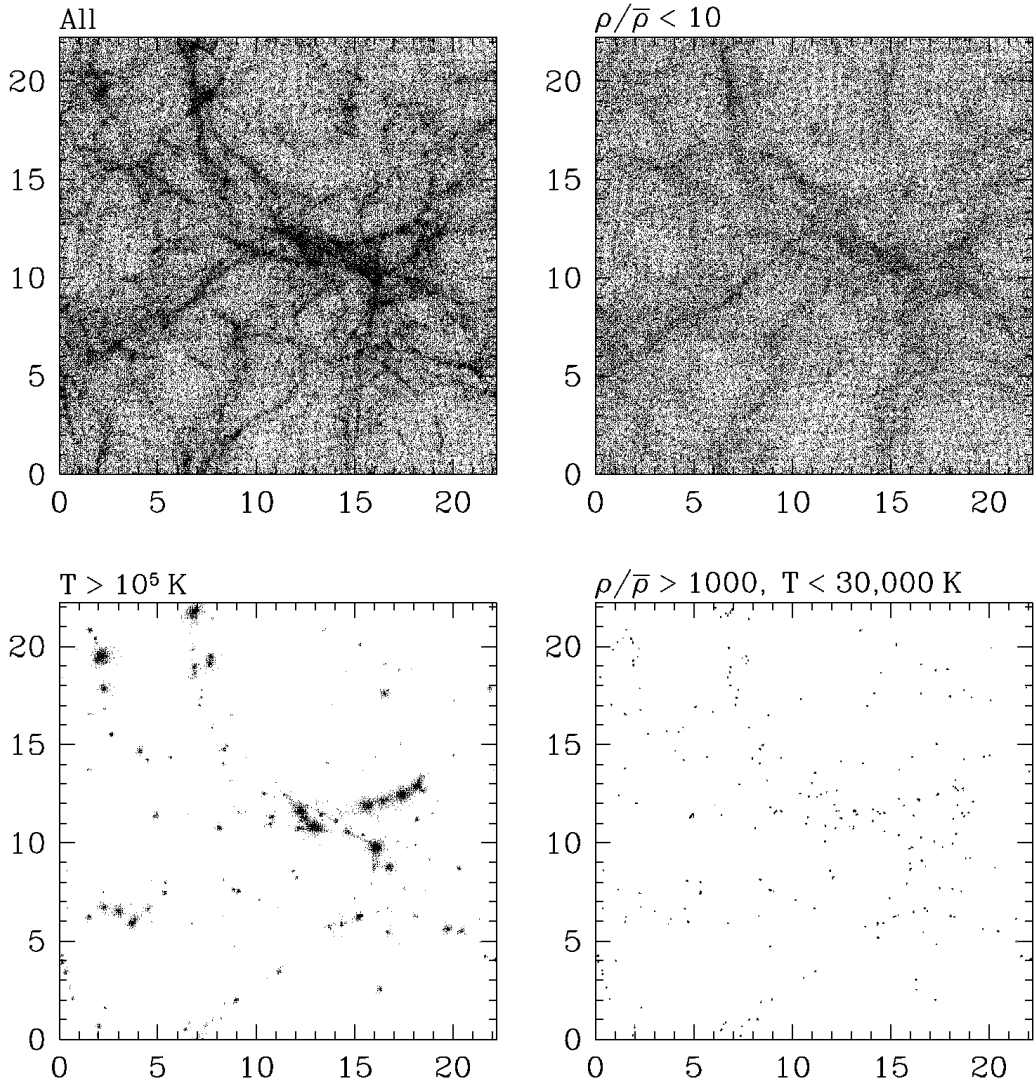}}
\caption{The distribution of gas particles in an SPH simulation of
the SCDM model, at $z=3$.  The simulation volume is a periodic cube,
22 comoving Mpc on a side, roughly 1/8 the comoving volume of the
most highly zoomed panel in Fig.~\ref{figCdm}.
The upper left panel shows a projection
of the full gas particle distribution.  The other panels show gas in
different overdensity and temperature regimes, as indicated.
}
\label{figSph}
\end{figure}

The upper left panel of Figure~\ref{figSph} shows the distribution
of gas particles in an SPH simulation
of the SCDM model at $z=3$.  The simulation volume is a periodic
cube 22 Mpc on a side (comoving, for $h=0.5$).
The pattern of filaments and voids is reminiscent of the large scale
structure in Figure~\ref{figCdm}, again reflecting the impact of non-linear
gravitational evolution on Gaussian initial conditions.
However, the extended features are much smaller than
those in Figure~\ref{figCdm}
because at $z=3$ the structure on larger scales has not yet collapsed.
The force and mass resolution are much higher than those of the
N-body simulation shown in Figure~\ref{figCdm}, but the price
of this increased resolution is a much smaller box size.

\begin{figure}
\centering
\centerline{\epsfxsize=5.0truein \epsfbox[85 450 535 685]{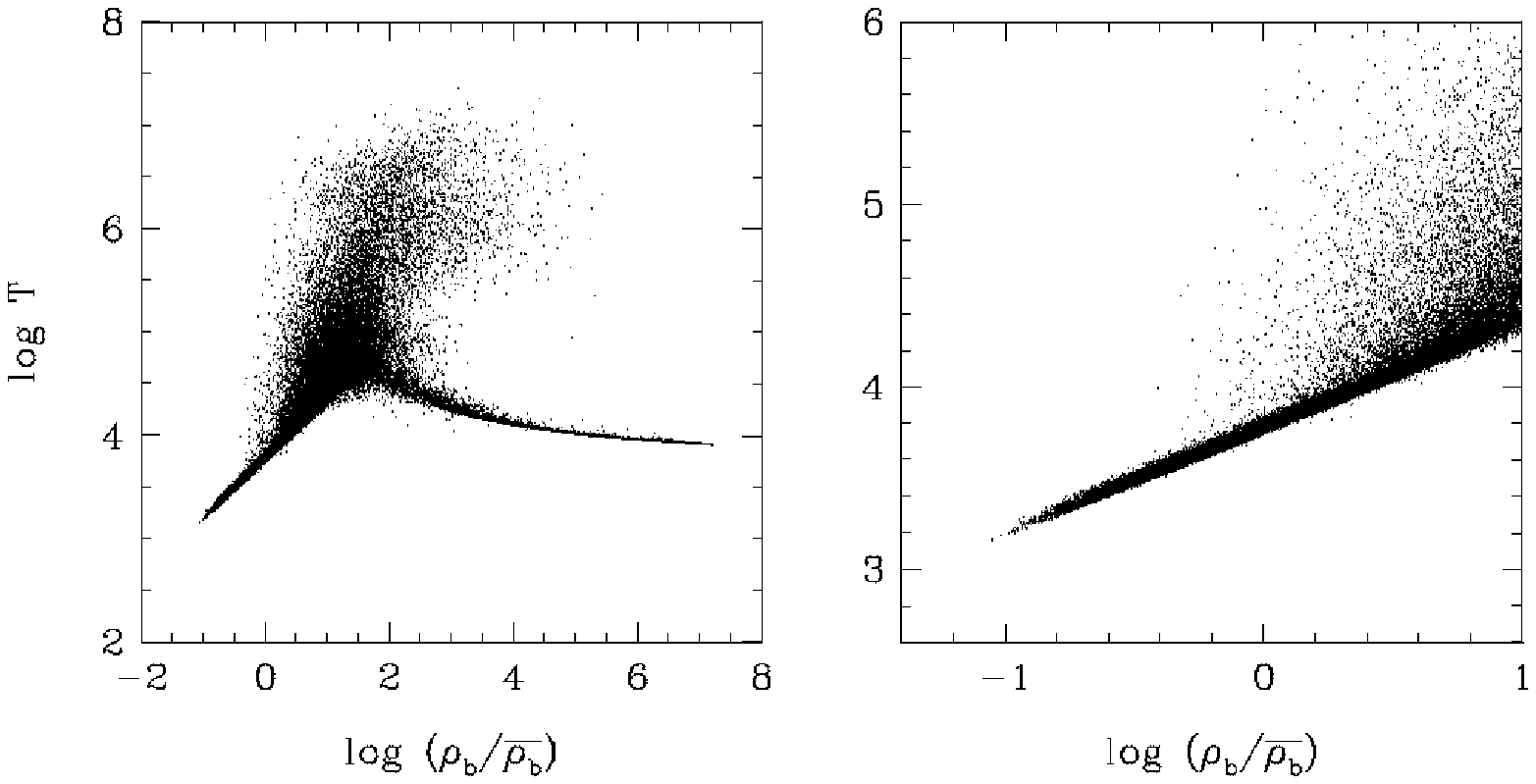}}
\caption{The distribution of gas particles in the temperature-density
plane, for the simulation shown in Fig.~\ref{figSph}.  
The right hand panel shows the temperature-density correlation
of the low density gas in greater detail.
}
\label{figRhoT}
\end{figure}

The left panel of Figure~\ref{figRhoT} shows the distribution of
the simulation's SPH particles in the temperature-density plane.
This plot reveals three distinct components to the gas distribution:
low density gas, with $\overden \la 10$ and typical temperature 
$T \la 10^5\;\K$, shock-heated gas, with typical overdensity
$\overden \sim 10-10^4$ and $T \sim 10^5-10^7\;\K$, and dense, cold
gas, with $\overden \ga 1000$ and $T \sim 10^4\;\K$.  The spatial
distribution of these three components is shown in the remaining
three panels of Figure~\ref{figSph}.  The low density component resides
in rounded underdense regions and filamentary structures.
The shock-heated component lies mainly in virialized halos, with
mildly shocked gas in filaments.  The cold gas knots are usually embedded
within more extended hot gas halos, as in Figure~\ref{figCluster}.
The simulation incorporates an algorithm that steadily turns cold,
dense, Jeans-unstable gas into stars, as described by KWH.
Gas in the high density cores of the cold knots has been partly 
converted to stars.  The star formation algorithm also returns thermal 
energy from supernova feedback to the surrounding gas, 
though because this gas is dense, the energy
is usually radiated away rapidly.  
As in Figure~\ref{figCluster}, the knots in Figure~\ref{figSph}
are visually inconspicuous because of their extreme
overdensity, but they contain about 5\% ($\sim 15,000$) of the
SPH particles.  

\begin{figure}
\centering
\centerline{\epsfxsize=3.5truein \epsfbox[50 245 570 660]{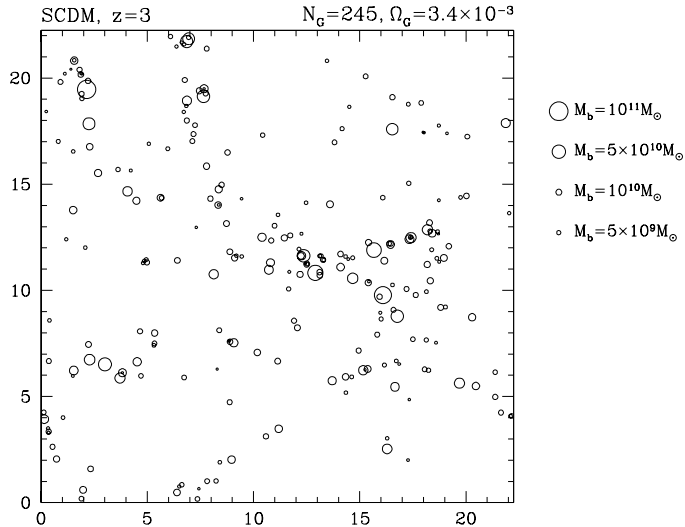}}
\caption{The distribution of ``galaxies'' in the SCDM simulation 
at $z=3$.  Each galaxy is represented by a circle with area proportional
to its baryonic mass (stars plus cold gas).
At $z=3$, the angular size of this comoving box is 12.7 arc-minutes,
and the redshift depth is $\Delta z = 0.0296$.
}
\label{figGal}
\end{figure}

In this Section, we focus on the cold, dense gas component and its
associated stars.  Because the knots of cold gas particles and
star particles are compact and distinct, it is easy to group them into 
simulated ``galaxies,'' as described in KWH.
Figure~\ref{figGal} shows the spatial distribution of galaxies at $z=3$. 
Each galaxy is represented by a circle whose area is proportional to its
baryonic mass (stars plus cold gas).  The more massive galaxies are 
predominantly stellar and the less massive galaxies more gas rich, though this
trend could be partly an artifact of the simulation's limited resolution,
which affects small galaxies more than larger ones.
By $z=3$, the largest galaxies in this simulation already have 
baryonic masses $M_b \sim 10^{11} M_\odot$.  
Comparison of Figure~\ref{figGal} and Figure~\ref{figSph} shows that
the galaxies approximately trace the structure in the underlying density field.
The spatial distribution and baryonic masses of the simulated galaxies
are not sensitive to the assumptions about star formation and supernova 
feedback within the range that we have tested.  Indeed, we get a 
nearly identical population of galaxies if we do not incorporate 
star formation and feedback at all and simply identify the galaxies
directly from the cold, dense gas (KWH).

\begin{figure}
\centering
\centerline{\epsfxsize=5.0truein \epsfbox[55 490 535 720]{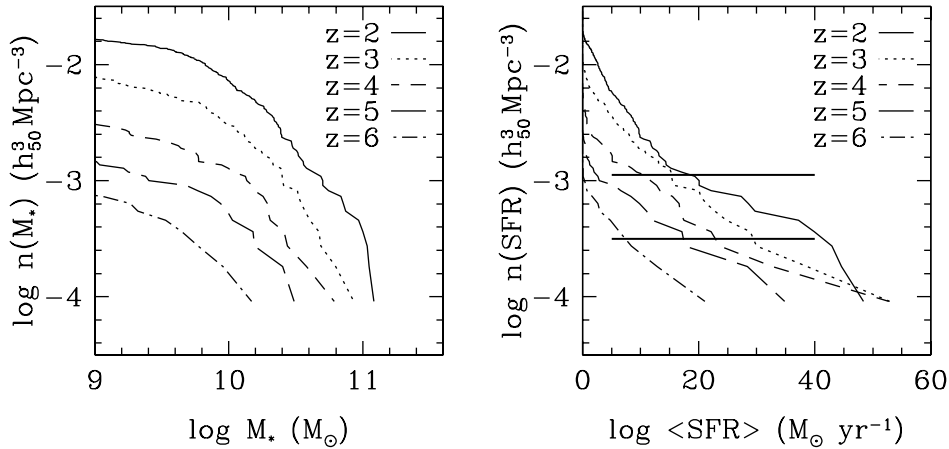}}
\caption{{\it Left}: Cumulative stellar mass function in the SCDM simulation
at the indicated redshifts: $n(M_*)$ is the comoving number density
of galaxies whose stellar mass exceeds $M_*$.
{\it Right:} Cumulative ``star formation rate'' function: 
$n({\rm SFR})$ is the comoving number density of galaxies 
whose time-averaged star formation rate exceeds $\langle {\rm SFR} \rangle$.
Horizontal lines indicate the space densities of the objects detected
by Steidel et al. (1996) and Lowenthal et al. (1997).  }
\label{figMassfun}
\end{figure}

The left panel of Figure~\ref{figMassfun} shows the cumulative stellar
mass function of the simulated galaxy population at $z=2$, 3, 4, 5, and 6.
The mass function rises steadily towards lower redshift as more gas
condenses into galaxies and small galaxies merge into larger systems.
Even at $z=6$, there are galaxies in the 
simulation volume with $M_* > 10^{10} M_\odot$.

Optical-band observations measure the rest-frame UV luminosities
of high redshift galaxies, so they respond mainly to the instantaneous
formation rate of massive stars rather than the accumulated stellar mass.
We cannot accurately estimate the instantaneous star formation 
rates of our individual simulated galaxies, but we can compute
a time-averaged star formation rate for each galaxy by dividing
its stellar mass by the age of the universe at redshift $z$.
The right panel of Figure~\ref{figMassfun} shows the cumulative
``star formation rate function,'' the comoving space density of
galaxies whose time-averaged star formation rate exceeds a
specified value.  The curves for different redshifts lie closer
together than the corresponding stellar mass function curves because
the age of the universe is smaller at high redshift.

Horizontal lines in Figure~\ref{figMassfun} mark the space densities
of objects found in two recent observational studies of 
Lyman-break galaxies, by Steidel et al.\ (1996) and Lowenthal et al.\ (1997).
If we assume, for example, that Steidel et al.\ (1996) detected
the galaxies with the highest star formation rates at $z \approx 3$,
then we see from the intersection of the dotted curve and the
lower horizontal line that the simulation predicts time-averaged
star formation rates 
$\langle {\rm SFR} \rangle \ga 30 M_\odot\,{\rm yr}^{-1}$
in these objects, and 
$\langle {\rm SFR} \rangle \sim 15 M_\odot\,{\rm yr}^{-1}$
in the fainter objects found by Lowenthal et al.\ (1997).
Steidel et al.\ (1996) estimate star formation rates of
$4 - 25 M_\odot\,{\rm yr}^{-1}$ (with $\Omega=1$, $h=0.5$)
for the galaxies in their sample, while Lowenthal et al.\ (1997)
estimate $3 - 8 M_\odot\,{\rm yr}^{-1}$ in their sample.
The conversion from observed UV luminosity to total star formation rate
depends on the adopted IMF and on corrections for dust extinction
(Steidel et al.'s estimates assume no dust correction and Lowenthal et al.'s
a modest correction), so the discrepancy between the predicted and
observed star formation rates may not be significant.
It certainly appears that the SCDM model has no difficulty
in producing a healthy population of high redshift galaxies.

From an observational perspective, the most encouraging message
of Figure~\ref{figMassfun} is 
that current studies have just scratched 
the luminous surface of the high-$z$ galaxy population.
At $z=3$, the simulation predicts a steep rise in the number of objects 
with decreasing luminosity, as a comparison of the 
Lowenthal et al.\ (1997) results to the Steidel et al.\ (1996) results
already suggests.
The simulation also predicts that deeper observations 
should find a significant population of galaxies at $z=6$ and beyond.

\begin{figure}
\centering
\centerline{\epsfxsize=3.5truein \epsfbox[95 410 465 720]{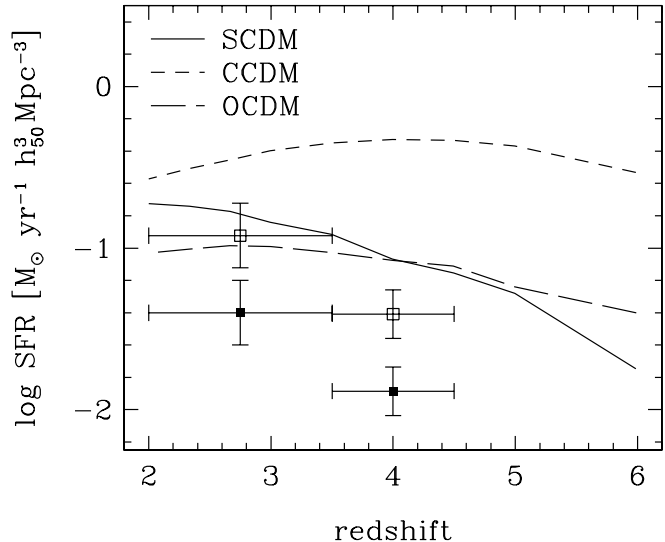}}
\caption{Global comoving density of star formation as a function of redshift
in the SCDM, CCDM, and OCDM models.  Filled squares show estimates 
based on the Hubble Deep Field from Madau (1997).  Open squares show
the same values with the dust extinction corrections suggested
by Pettini et al.\ (these proceedings).
}
\label{figSfr}
\end{figure}

Figure~\ref{figSfr} shows the globally averaged, instantaneous star
formation rate from $z=6$ to $z=2$.  In addition to the SCDM model
discussed above, we show results for a COBE-normalized, $\Omega=1$
CDM model (CCDM) and an open-universe CDM model with
$\Omega_0=0.4$ (OCDM, also COBE-normalized; for details of the
models and simulations see \cite{cwkh97b}).
The CCDM model has a higher amplitude of mass fluctuations
than the other two models, and it therefore produces more massive
galaxies and forms them earlier.
By $z=3$ the global star formation rate in this model
has already hit its peak and begun to decline.
The star formation history of the OCDM model is similar to that
of SCDM but shifted towards slightly
higher redshift, as expected from the difference in fluctuation
growth rates between an $\Omega=1$ universe (where clustering
grows steadily with expansion factor) and an open universe (where the
growth of clustering ``freezes out'' towards low redshift).

With our adopted prescription for star formation, the rate at which
a galaxy forms stars is essentially governed by the rate at which
halo gas cools and settles into the central object.
There is no requirement of a ``triggering'' event (such as an interaction
with another galaxy) to initiate star formation, and supernova feedback, 
though incorporated in the algorithm, does not have a dramatic
effect on the surrounding medium until the cold gas supply begins
to run out.  Given these assumptions about star formation physics,
the numbers in Figure~\ref{figSfr} should be regarded as lower limits
to the true predictions of these cosmological models, since
the simulations only resolve galaxies with circular velocities
$v_c \ga 100\;{\rm km}\;{\rm s}^{-1}$.  
Increasing the resolution of the simulations would shift the
global star formation rates upward, especially at higher redshift,
where massive halos are comparatively rarer.  

The solid squares in Figure~\ref{figSfr} show the estimates of
the global star formation rate from Madau (1997), based on the
Hubble Deep Field.  The open squares are shifted upwards by a 
factor of three, the correction for dust extinction suggested
by Pettini et al.\ (these proceedings).
The simulations again predict more high redshift star formation
than is inferred from the observations.  Pettini et al.'s extinction
correction removes most of the discrepancy, but the predicted 
star formation rates would be larger for higher resolution simulations
(as discussed above) or for the larger baryon density
advocated by Rauch et al.\ (1997b; see discussion in \S 5 below).
Our simulations predict somewhat higher star formation rates and 
stellar mass functions than semi-analytic calculations of galaxy
formation with the same underlying cosmology (\cite{baugh97}).  
We are still investigating the source of this difference, but
it appears to result mainly from the strong feedback invoked in
the semi-analytic calculations, which suppresses star formation
in low and intermediate mass halos (Baugh, private communication).
This strong feedback is incorporated in the semi-analytic calculations
mainly in order to match the faint end of the galaxy luminosity
function at $z=0$ (\cite{kauffmann93}; \cite{cole94}); the suppression
of star formation in high redshift galaxies is an indirect consequence.
The inference of the global star formation rate from observations
depends on the assumed extinction correction and on the assumed IMF,
so it is probably premature to say whether the numerical or semi-analytic
predictions are in better agreement with the data.
However, Madau (these proceedings) shows that the extinction and IMF
assumptions can be tested internally by predicting the luminosity densities
in different bands at various redshifts.
Combinations of this work, semi-analytic calculations, hydrodynamic
simulations, and further observations should go a long way to
unraveling the physics of the high redshift galaxy population,
in particular indicating whether supernova feedback or some similar
mechanism suppresses gas cooling and star formation in high redshift halos.

\section{Photoionization and the Lyman-alpha forest}

We now turn our attention from galaxies to the diffuse intergalactic
medium (IGM), the low density gas component shown in the upper right
panel of Figure~\ref{figSph}.  In addition to gravity, the crucial
physics that a simulation must incorporate in order to model this
component correctly is photoionization by the UV bacgkround radiation.
The SCDM simulation in Figure~\ref{figSph} incorporates the UV
background field computed by Haardt \& Madau (1996), based on the
UV output of the observed quasar population.  (Star forming galaxies
could also contribute to the UV background if ionizing photons
escape from the interstellar medium.)  Because the recombination
time at low density is long, this UV background keeps the diffuse
IGM (i.e., the $\overden \la 10$ component) highly photoionized,
with a typical hydrogen neutral fraction $\sim 10^{-6} - 10^{-4}$.

The right hand panel of Figure~\ref{figRhoT} shows an expanded view
of the temperature-density diagram for this diffuse IGM component.
A small fraction of the gas with $\overden < 10$ has been shock heated
to $T > 10^{4.5}\;\K$, but most of it follows a tight temperature-density
relation, $T = T_0(\overden)^\gamma$, with $T_0 \approx 6000\;\K$
and $\gamma \approx 0.6$.  This relation emerges from the interplay
between photoionization heating --- stronger at higher density because
of the higher neutral fraction --- and adiabatic cooling, caused by the
expansion of the universe.  The density of the gas is too low
for significant free-free or recombination cooling.  The precise values of
$T_0$ and $\gamma$ depend on redshift, on the baryon density
$\Omega_b h^2$, on the UV background spectrum, and on the history
of hydrogen and helium reionization. However, the existence of a tight
temperature-density relation is a quite general consequence of the 
simple processes, photoionization heating and adiabatic cooling, which
determine the thermal state of this diffuse gas.
At $z \sim 2-4$ the parameters of the relation generally lie in
the range $4000 \la T_0 \la 15,000\;\K$ and $0.3 \la \gamma \la 0.6$,
for reasonable assumptions about the baryon density and reionization
history (\cite{hg97}).

\begin{figure}
\centering
\centerline{\epsfxsize= 5.3 truein \epsfbox[45 435 550 720]{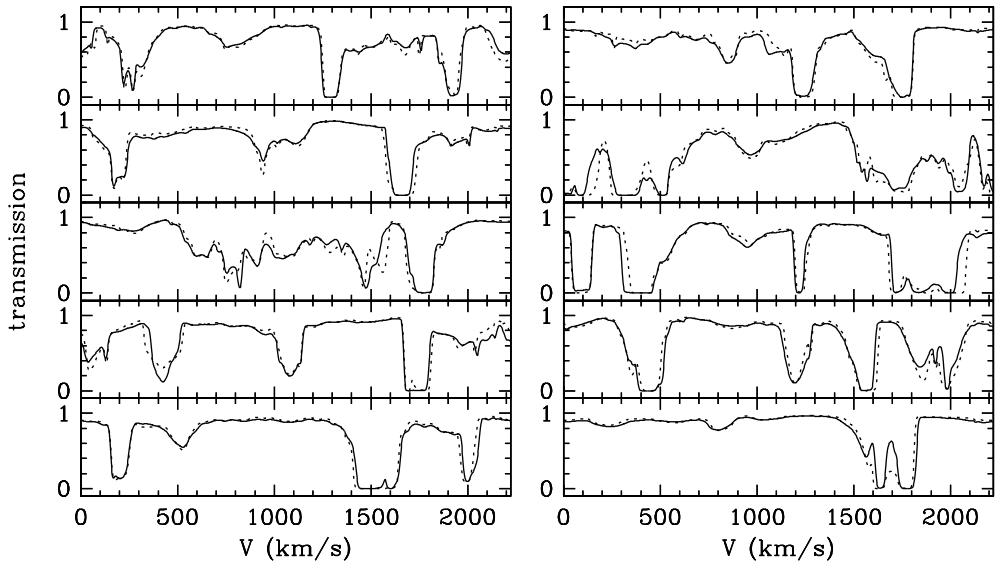}}
\caption{\lya absorption spectra along ten random lines of sight
through the SCDM simulation box at $z=3$ (solid lines).  The transmitted flux
$e^{-\tau}$ is plotted against velocity; the corresponding observed
wavelength would be $\lambda = \lambda_\alpha (1+z) (1+v/c)$.
Dotted lines show spectra along the same lines of sight computed by
the approximate N-body technique described in the text.
}
\label{figSpec}
\end{figure}

Resonant \lya absorption from even a small amount of neutral hydrogen
can significantly depress the UV continuum from a background quasar
(\cite{gunn65}).  Figure~\ref{figSpec} shows the absorption that
would be produced in the spectra of background quasars along
ten random lines of sight through the simulation box.
Neutral atoms at velocity $v$ (relative to the leading edge of the box)
produce absorption at observed wavelength 
$\lambda = \lambda_\alpha (1+z) (1+v/c)$, where $z=3$ and
$\lambda_\alpha = 1216\AA$ is the rest wavelength of Ly$\alpha$.
As light passes through the box, the spatially varying HI density
field imprints an irregular picket fence pattern of absorption ``lines,''
the \lya forest (\cite{lynds71}; \cite{sargent80}).

This natural explanation for the \lya forest is probably the most
remarkable result to emerge from hydrodynamic cosmological
simulations (\cite{cen94}; \cite{zhang95}; \cite{hkwm96}).
The underlying cosmological models (e.g., SCDM)
are motivated by inflationary cosmology and by analyses of CMB anisotropies
and large scale structure. 
Starting with such a model, one 
puts in gravity and photoionization,
and the \lya forest falls out ``for free,'' with no special
tweaks.  Saturated absorption lines, with 
neutral column density $\Nh \ga 10^{14.5}\;\cdunits$,
typically arise when the line of sight crosses one of the filaments
or sheets clearly visible in Figure~\ref{figSph}.  
More common, optically thin lines are usually associated with
milder density fluctuations in gas fairly close to the cosmic
mean density.  The weakest lines often correspond to local density
maxima in underdense regions.

Most \lya forest lines with $\Nh \ga 10^{14.5}\;\cdunits$ are known to
have associated metal lines (\cite{meyer87}; \cite{cowie95}; 
\cite{songaila96}; \cite{womble96}),
so this scenario for the \lya forest requires enrichment of the
diffuse IGM by early star formation.  
Enrichment to the required level appears plausible on
theoretical grounds (\cite{go97}; \cite{haiman97}; \cite{miralda97}),
and simulations that assume an IGM metallicity $Z \sim 10^{-2.5} Z_\odot$
account for the observed metal lines fairly well
(\cite{haehnelt96}; \cite{hdhwk97}ab; \cite{rhs97}).
Simulations of the SCDM model also reproduce the observed incidence
of high column density, damped \lya absorption
($\Nh \geq 10^{20.3}\;\cdunits$),
which arises in the
cold, dense gas of the high redshift galaxies discussed earlier
(\cite{kwhm96}; \cite{gkhw97}).
The complex kinematics revealed by the metal lines in high 
column density systems supports the picture of hierarchical galaxy
formation (\cite{haehnelt96}, 1997; \cite{rhs97}).

Physical properties of the low column density absorbers
($\Nh \la 10^{15}\;\cdunits$) in cosmological
simulations are quite different from those of the ``clouds'' envisioned
in many earlier models of the \lya forest.  Because of the low densities,
the absorbing systems are far from dynamical or thermal equilibrium.
The neutral fractions and neutral hydrogen densities are very low,
and the systems must therefore be quite large ($\sim 100$ kpc) in 
order to have HI column densities $\sim 10^{14}\;\cdunits.$
This large scale is implied by observations of quasar pairs,
independently of the simulations themselves 
(\cite{bechtold94}; \cite{dinshaw94}, 1995; \cite{crotts97}).
Hubble expansion across the spatially extended absorber
is often the dominant contributor to the absorption line velocity width,
even though the absorbing region may be shrinking in {\it comoving}
coordinates.  In models where \lya forest lines are produced by dense, compact
clouds, the line widths are instead determined by thermal or ``turbulent''
broadening, and the wings of the lines are caused by high velocity atoms.
In the cosmological simulations picture, the line wings show the
absorbing structure itself fading into the background, like mountains
into foothills.

Gunn \& Peterson (1965) showed that \lya absorption by a uniform IGM
with neutral hydrogen density $\nh$ would depress the UV continuum of
a quasar blueward of \lya by a factor $e^{-\tau_{\rm GP}}$, where
\begin{equation}
\tau_{\rm GP} = \frac{\pi e^2}{m_e c}\; f_\alpha \lambda_\alpha
H^{-1}(z) \nh.
\label{taugp}
\end{equation}
Here $f_\alpha$ is the \lya oscillator strength, and $H(z)$ is the Hubble
parameter at redshift $z$.
In cosmological simulations, there is no sharp distinction between the
lines and the background, and one can interpret the \lya forest itself
as a fluctuating Gunn-Peterson effect (\cite{hkwm96}; \cite{miralda96};
\cite{cwkh97b}).  This picture of the \lya forest bears some resemblance
to the ``intracluster medium'' view of quasar absorption lines suggested by
Bahcall \& Salpeter (1965), and it is quite similar to the inhomogeneous
IGM models described by McGill (1990) and Bi (1993).

What turns the fluctuating Gunn-Peterson idea from a novelty into
a powerful conceptual tool is the simplicity of the physics that
governs the ionization state of the low density gas.
This gas is in photoionization equilibrium, so the neutral hydrogen
density $\nh \propto \rho^2 T^{-0.7}/\Gamma$, where $\Gamma$ is
the photoionization rate and the $T^{-0.7}$ factor accounts for the
temperature dependence of the hydrogen recombination coefficient near
$T \sim 10^4\;\K$.
The low density gas responsible for most of the \lya absorption
follows the tight temperature-density relation discussed above,
$T = T_0 (\overden)^\gamma$.
From the Gunn-Peterson formula~(\ref{taugp}), we therefore 
expect $\tau \propto \nh \propto (\overden)^\beta$,
where $\beta = 2-0.7\gamma \approx 1.6$.
In principle, $\overden$ here refers to the {\it gas} overdensity,
but because the temperature is low, pressure gradients are small
compared to gravitational forces, and the gas traces the dark matter
quite well.

Putting in constants, we arrive at an equation that we might
call the {\it fluctuating Gunn-Peterson approximation},
\begin{eqnarray}
\tau(\lambda_{\rm obs}) = &
 0.172 \left(\frac{\rho}{\overline \rho}\right)^\beta
\left(1 + \frac{dV_{\rm los}}{H(z) dx}\right)^{-1}
\left(\frac{1+z}{4}\right)^6
\left(\frac{H(z)/H_0}{5.51}\right)^{-1} h^{-1} \;\times & \nonumber \\
& \left(\frac{\Omega_b h^2}{0.0125}\right)^2 
\left(\frac{T_0}{10^4\;{\rm K}}\right)^{-0.7}
\left(\frac{\Gamma}{10^{-12}\;{\rm s}^{-1}}\right)^{-1}\;, & \label{fgpa} 
\end{eqnarray}
where $\overden$ is the overdensity at the position where the
redshift (cosmological plus peculiar velocity) is 
$\lambda_{\rm obs}/\lambda_\alpha - 1$ and $dV_{\rm los}/dx$
is the derivative of the line-of-sight peculiar velocity at the same
position.  The peculiar velocity term accounts for the mapping from
real space to redshift space.  
A cosmological model determines the statistical properties of the
density and peculiar velocity fields, and the values of $\Omega_0$ 
and the cosmological constant determine the Hubble parameter ratio
\begin{equation}
H(z)/H_0 = \left[\Omega_0(1+z)^3 + (1-\Omega_0-\lambda_0)(1+z)^2 + 
\lambda_0\right]^{1/2}. 
\label{hratio}
\end{equation}
Here $\lambda_0$ is the cosmological constant divided by 
$3H_0^2$ (so that a spatially flat universe has $\Omega_0+\lambda_0=1$).
The fiducial value of $H(z)/H_0$
in equation~(\ref{fgpa}) is computed at $z=3$
for $\Omega_0=0.3$, $\lambda_0=0$.  

Equation~(\ref{fgpa}) is valid if all gas lies on the temperature-density
relation and thermal broadening and collisional ionization can be
ignored.  The approximation breaks down when $\overden \ga 10$,
but these regions occupy a small fraction of the spectrum.
The reasoning that leads to equation~(\ref{fgpa}) also underlies
semi-analytic models of the \lya forest developed by
Bi, Ge, \& Fang (1995), Bi \& Davidsen (1997), 
and Hui, Gnedin, \& Zhang (1997), which use the lognormal or truncated
Zel'dovich approximation to compute density and peculiar velocity
fields (these approximations also incorporate thermal broadening).
For a more accurate, ``semi-numerical'' approach that is much
cheaper than a high resolution hydrodynamic calculation, one can
run a lower resolution N-body simulation, compute the density
field from the particle distribution, impose the temperature-density
relation, and extract spectra.  This technique uses a fully non-linear
solution for the density and velocity fields, but it still assumes
that gas traces dark matter and that all gas lies on the temperature-density
relation, approximations that are good but not perfect.
The dotted lines in Figure~\ref{figSpec} show spectra along the same
ten lines of sight computed by this method from an N-body simulation
with the same initial conditions.  Agreement with the full hydrodynamic
simulation is very good over most of the spectrum, breaking down 
in a few regions where shock heating pushes gas off the 
assumed temperature-density relation.  
Similar techniques have been used by Gnedin \& Hui (1997), who
also incorporate an approximate treatment of gas pressure in the
N-body calculation, and by Petitjean, M\"ucket, \& Kates (1995)
and M\"ucket et al.\ (1996), who use a more elaborate method
to compute gas temperatures.

One can simplify equation~(\ref{fgpa}) by ignoring the peculiar
velocity derivative $dV_{\rm los}/dx$.
The resulting approximation is somewhat
less accurate, but it is more analytically tractable because it
implies a one-to-one relation between the optical
depth and the local overdensity, $\tau = A (\overden)^\beta$.
Peculiar velocities affect $\tau$ and the redshift space density
in the same sense --- a converging flow enhances both, for example ---
but the redshift space $\tau$--$\rho$ relation has larger scatter than
the real space $\tau$--$\rho$ relation because the neutral hydrogen
fraction depends on the real space, physical density.
Nonetheless, spectra from the hydrodynamical simulations,
which include peculiar velocities, thermal broadening, shock heating,
and collisional ionization, exhibit an impressively tight 
correlation between $\tau$ and $\rho$ even in redshift space
(\cite{cwkh97b}, figure 6).

One can use this simpler form of the fluctuating Gunn-Peterson
approximation to estimate the density parameter of gas in the
\lya forest.  A directly observable property of \lya forest spectra
is the optical depth distribution $P(\tau)$, where $P(\tau)d\tau$
is the fraction of pixels with optical depth in the range 
$\tau \longrightarrow \tau+d\tau$.  With a one-to-one mapping
between density and optical depth, the implied mean density
is ${\overline \rho} = \int_0^\infty \rho(\tau) P(\tau) d\tau$.
Using equation~(\ref{fgpa}) with $dV_{\rm los}/dx=0$ to obtain 
$\rho(\tau)$, one finds
\begin{eqnarray}
\Omega_{\rm WIGM} \equiv f_{\rm WIGM} \Omega_b
= & 0.021 h^{-3/2} 
\left(\frac{
\left[\int_0^\infty \tau^{1/\beta} P(\tau) d\tau\right]^{\beta/2}}
{0.70} \right) \left(\frac{4}{1+z}\right)^3 \;\times & \nonumber \\
& \left(\frac{H(z)/H_0}{5.51}\right)^{1/2} 
\left(\frac{T_0}{10^4\;\K}\right)^{0.35} 
\left(\frac{\Gamma}{10^{-12}\; {\rm s}^{-1}}\right)^{1/2}\;. & \label{igm}
\end{eqnarray}
Further details of this calculation are given by Weinberg et al.~(1997);
equation~(\ref{igm}) is equivalent to equation~(22) of that paper.
Here $f_{\rm WIGM}$ is the fraction of baryons in the ``warm''
IGM that produces most of the \lya forest, more specifically, the
fraction of gas that has not been substantially shock heated and
therefore lies on the temperature-density relation.
Roughly speaking, this is gas with $\overden \la 10$.

As a fiducial value of the optical depth integral in equation~(\ref{igm}),
we have used the
value 0.70 obtained from spectra of the SCDM model at $z=3$.
The $P(\tau)$ computed from these spectra with artificial noise
and other instrumental effects is an almost perfect match 
to the $P(\tau)$ obtained from Keck HIRES spectra by
Rauch et al.\ (1997b; see their figure~4).  We therefore use $P(\tau)$
from the noiseless simulated spectra as an estimate of the true
optical depth distribution, free of observational artifacts.
Since $\tau$ cannot be determined accurately in saturated regions,
we set $\tau=3$ whenever $\tau > 3$ (9\% of the spectrum), so that
$\Omega_{\rm WIGM}$ is slightly underestimated.
Ignoring peculiar velocities also causes equation~(\ref{igm}) to
underestimate $\Omega_{\rm WIGM}$, but only by $\sim 5-10\%$
(see \cite{wmhk97}).
Cosmological parameters influence the estimate of $\Omega_{\rm WIGM}$
only through $h$ and the ratio $H(z)/H_0$.
At $z=3$, this ratio is 5.51 for for $\Omega_0=0.3$, $\lambda_0=0$,  
4.46 for $\Omega_0=0.3$, $\lambda_0=0.7$, and 8.0 for $\Omega_0=1$,
$\lambda=0$.

With a conservative value for the IGM temperature, $T_0=5000\;\K$,
and a conservative constraint on the photoionization rate, 
$\Gamma \geq 7 \times 10^{-13}\;{\rm s}^{-1}$ (see \cite{rauch97}),
equation~(\ref{igm}) implies a lower bound
$\Omega_{\rm WIGM} \geq 0.014 h^{-3/2}$.
This is a substantial fraction of the total baryon density 
$\Omega_b=0.02h^{-2}$ derived from big bang nucleosynthesis with
the Burles \& Tytler (1997) estimate of the primordial deuterium
abundance (see also \cite{tytler96}, who estimate $\Omega_b=0.024h^{-2}$,
and \cite{rugers96}ab, who estimate a much smaller density, 
$\Omega_b = 0.006 h^{-2}$).
Comparison of these values shows that the warm IGM is the dominant reservoir
of high redshift baryons, with 
$f_{\rm WIGM} =\Omega_{\rm WIGM}/\Omega_b \geq 0.7 h^{1/2}$.
This conclusion does not rely on the simulations explicitly,
only on the Rauch et al.\ (1997b) $P(\tau)$ measurement
and on the assumption that \lya 
forest absorbers are, on average, no more extended in redshift
space than they are in real space (\cite{wmhk97}).
However, the simulations do {\it predict} that most baryons reside
in the warm IGM at $z=3$, in accord with this observational inference.
The values of $f_{\rm WIGM}$ (specifically, the fraction of the baryons
with $\overden < 10$) are 0.69, 0.67, and 0.49 for the SCDM, OCDM,
and CCDM models, respectively.
One can use equation~(\ref{igm}) to estimate $\Omega_b$ by assuming
a value of $f_{\rm WIGM}$ and putting in best estimates for other
parameters.  For example, with $f_{\rm WIGM}=0.7$,
$\Gamma = 1.4 \times 10^{-12}\;{\rm s}^{-1}$ (\cite{haardt96}),
$T_0=6000\;\K$, and $H(z)/H_0=5.51$ at $z=3$, one obtains
$\Omega_b=0.03 h^{-3/2}$.
The largest uncertainty at present is the Rauch et al.\ (1997b)
$P(\tau)$ measurement itself, which is based on a fairly small
quasar sample.  However, the same analysis at $z=2$ (again with
the Rauch et al.\ data) yields nearly the same $\Omega_b$, and results
from a larger data sample should be available soon.

\section{Cosmology with the Lyman-alpha forest}

Most traditional methods of characterizing \lya forest spectra begin
by decomposing each spectrum into a set of lines, usually assumed
to have Voigt profiles.  (For a line with $\Nh \la 10^{19}\;\cdunits$,
damping wings are insignificant, and a Voigt profile is simply a
Gaussian optical depth profile.)
Comparisons between simulated spectra and observational data using
line decomposition methods show that CDM models with reasonable
parameters match the observed statistical distributions of neutral 
hydrogen column densities and line widths fairly well
(\cite{miralda96}; \cite{dhwk97}; \cite{gnedin97}; \cite{zhang97}).
However, while any spectrum can be represented as a superposition
of a sufficient number of Voigt-profile lines, the simulations suggest
that such a superposition is not a natural or robust description of
the \lya forest.  The individual ``absorbers'' in the simulations are
extended, out-of-equilibrium structures, which do not produce Voigt-profile 
lines.  Furthermore, because there are no clear boundaries between the
absorbers and the background medium, the mere division of the spectrum
into a collection of lines is somewhat artificial.

Equation~(\ref{fgpa}) suggests a more natural point of view: to a good
approximation, a \lya forest spectrum is a non-linear map of the 
continuous, redshift-space density field along the line of sight.
Statistical measures that treat the spectrum as a continuous field
are likely to be more powerful than line decomposition methods for
discriminating cosmological models, because they involve minimal manipulation
of the observational data and because they are better attuned
to the underlying physics of the absorbing medium.
One particularly simple measure is the distribution function of
flux decrements $P(D)$, where $D = 1-e^{-\tau}$ and $P(D)dD$ is the
fraction of pixels with flux decrement in the
range $D \longrightarrow D+dD$ (\cite{miralda96}; \cite{cwhk97};
\cite{kim97}; \cite{rauch97}).  This statistic is directly analogous
to the counts-in-cells distribution frequently used in studies
of large scale structure (\cite{saslaw89}; \cite{kim97a}).
Another simple statistic is the threshold crossing frequency,
the average number of times per unit redshift that the spectrum
crosses a specified transmission threshold 
(\cite{miralda96}; \cite{cwhk97}; \cite{kim97}).
A plot of this crossing frequency against spectral filling factor
is directly analogous to the ``genus curve'' used to characterize
the topology of the galaxy density field (\cite{gwm87}).

We are currently using simulations --- both full hydrodynamic calculations
and the ``pseudo-hydro''
N-body technique described earlier --- to see how these statistical
measures respond to properties of the cosmological model.
The preliminary results accord well with expectations from the
fluctuating Gunn-Peterson approximation and the analogies to large scale
structure statistics.  For models with Gaussian initial conditions,
the flux decrement distribution $P(D)$
depends mainly on the amplitude of the mass power spectrum on
comoving scales $\sim 1\hmpc$.  Increasing the power spectrum amplitude
skews the distribution of overdensities $P(\overden)$, 
and this in turn skews the
flux decrement distribution because of the correlation between
$\tau$ and $\overden$.  Models with non-Gaussian initial conditions
produce a different overdensity distribution (see, e.g., \cite{protogeros97}),
so we expect that $P(D)$ will also be a sensitive discriminant
between Gaussian and non-Gaussian models, though we have not yet studied
this issue numerically.
The threshold crossing statistic measures the ``corrugation'' of the
absorption spectrum --- a spectrum with more small scale structure
has more crossings per unit redshift.
Like its 3-dimensional cousin, the genus curve, the threshold
crossing frequency depends on the logarithmic {\it slope} of the
underlying mass power spectrum (\cite{doroshkevich70}; \cite{adler81};
\cite{bardeen86}; \cite{hgw86}).
A mass density field with a bluer power spectrum produces choppier
absorption and a higher crossing frequency.  
Thermal broadening and peculiar velocities also influence the crossing
frequency by physically smoothing the absorption on small scales.

One can investigate larger scale clustering by artificially smoothing
the absorption spectrum (e.g., convolving with a Gaussian filter)
and applying the same measures, just as one might study the galaxy
counts-in-cells distribution or the topology of the galaxy density
field at many spatial scales.
For example, if the mass distribution is strongly clustered on large
scales, then the flux decrement distribution $P(D)$ remains skewed
even in the smoothed spectrum.  If the large scale clustering is weak,
then $P(D)$ quickly settles towards a sharply peaked,
roughly Gaussian form as the smoothing is increased.
One can also probe larger scale clustering via the flux
correlation function (\cite{pr93}; \cite{zuo94}; \cite{miralda96})
or power spectrum (\cite{cwkh97c}).  These measures are closely
related to the correlation function and power spectrum of the mass
density itself, and by measuring them at several redshifts one can
observe the growth of density fluctuations over time.

High resolution spectrographs on large telescopes yield \lya forest
spectra of spectacular quality.  As the above discussion indicates,
the information content of these spectra is prodigious, and it can
be directly related to cosmological parameters and to basic properties
of the mass distribution at high redshift.
Equally important, the simplicity of the physical processes that
affect the low density IGM allows cosmological simulations to make 
robust quantitative predictions for a specified cosmological model.
The \lya forest therefore offers an excellent arena in which to test
theories of structure formation.

In addition to comparing specific models to observations, one 
can work backwards from the absorption spectra to derive the shape
and amplitude of the primordial mass power spectrum.
This approach is described by Croft et al.\ (1997c), and tests on
artificial spectra from the simulations show that it works remarkably
well, neatly circumventing the uncertain physics of galaxy formation,
which complicates the interpretation of power spectra measured
from galaxy redshift surveys.  Application to Songaila \& Cowie's (1996)
spectrum of Q1422+231 
yields a $z=3$ power spectrum consistent with the SCDM and OCDM
models considered here.  This cosmological test will become
much more sensitive in the near future, when the technique is 
applied to larger quasar samples.

\section{Final Remarks}

Taken together, the results described in the preceding sections tell an
encouraging tale: as cosmological simulations incorporate progressively
more realistic physics, they explain a 
progressively wider range of observational data.  
Collisionless N-body simulations show that non-linear gravitational evolution
transforms Gaussian primordial fluctuations
into a pattern of sheets, filaments, tunnels, and voids
that closely resembles the structure seen in 3-dimensional maps
of the galaxy distribution.  Simulations that add gas dynamics,
radiative cooling, and star formation produce dense clumps of stars and 
cold gas with sizes and masses similar to those of luminous galaxies.
Simulations that include photoionization by the UV radiation
background match the \lya forest absorption seen in the spectra of high
redshift quasars.  

Numerical simulations of structure formation in a universe
dominated by collisionless dark matter offer a unified account of
disparate observational phenomena, from large scale structure to
the \lya forest.  The gravitational instability scenario can also
account for important phenomena not covered in this review,
such as CMB anisotropies (Wright, these proceedings) and
hot gas in galaxy clusters (Donahue, these proceedings).
Crucial questions about the nature and amount of dark matter,
the geometry of spacetime, and the origin of primordial fluctuations
remain open, but the range of plausible answers already seems
much narrower than it did 10 years ago.  Space observatories,
large telescopes, and ambitious surveys are providing a rich, diverse
harvest of data, and cosmological simulations are becoming more powerful
and sophisticated, so we can expect great progress in the decade ahead.

\acknowledgments
We are grateful to our numerous collaborators for stimulating conversations
on these topics, especially Rupert Croft, Romeel Dav\'e, 
Uffe Hellsten, and Jordi Miralda-Escud\'e.  We acknowledge research support 
from NASA and the NSF and computing support from the San Diego, Pittsburgh,
and Illinois supercomputer centers.


\begin{thebibliography}
{\parskip=0pt
\bibitem[Adler 1981]{adler81}
Adler, R. J. 1981, The Geometry of Random Fields (Wiley: New York)

\bibitem[Bahcall \& Salpeter 1965]{bahcall65}
Bahcall, J. N., \& Salpeter, E. E. 1965, \apj, 142, 1677

\bibitem[Bardeen et al.\ 1986]{bardeen86}
Bardeen, J., Bond, J. R., Kaiser, N., \& Szalay, A. 1986,
\apj, 304, 15

\bibitem[Barnes \& Hut 1986]{barnes86}
Barnes, J.E., \& Hut, P. 1986, Nature, 324, 446

\bibitem[Baugh et al.\ 1997]{baugh97}
Baugh, C. M., Cole, S., Frenk, C. S., \& Lacey, C. G. 1997, \apj, submitted
(astro-ph/9703111)

\bibitem[{Bechtold et al.\ }1994]{bechtold94}
Bechtold, J., Crotts, A. P. S., Duncan, R. C., \& Fang, Y. 1994, \apj, 437, L83

\bibitem[Bhavsar, Aarseth, \& Gott 1981]{bhavsar81}
Bhavsar, S. P., Aarseth, S. J., \& Gott, J. R. 1981, \apj, 246, 656

\bibitem[{Bi }1993]{bi93}
Bi, H.G., 1993, \apj, 405, 479

\bibitem[{Bi \& Davidsen }1997]{bi97}
Bi, H.G., \& Davidsen, A. 1997, \apj , 479, 523

\bibitem[{Bi, Ge, \& Fang }1995]{bi95}
Bi, H., Ge, J., \& Fang, L.-Z. 1995, \apj, 452, 90

\bibitem[Burles \& Tytler 1997]{burles97}
Burles, S., \& Tytler, D. 1997, \aj, submitted

\bibitem[Cen et al.\ 1994]{cen94}
Cen, R., Miralda-Escud\'e, J., Ostriker, J.P., \& Rauch, M. 1994, \apj, 437, L9

\bibitem[Centrella \& Melott 1983]{centrella83}
Centrella, J., \& Melott, A. L. 1983, Nature, 305, 196

\bibitem[Cole et al.\ 1994]{cole94}
Cole, S., Aragon-Salamanca, A., Frenk, C. S., Navarro, J. F., \& Zepf, S. E.
1994, \mnras, 271, 781

\bibitem[Cole et al.\ 1997]{cole97}
Cole, S., Weinberg, D. H., Frenk, C. S., \& Ratra, B. 1997,
\mnras, 289, 37

\bibitem[Cowie et al.\ 1995]{cowie95}
Cowie, L.L., Songaila, A., Kim, T.-S., \& Hu, E.M. 1995, \aj, 109, 1522

\bibitem[Croft et al.\ 1997a]{cwhk97}
Croft, R. A. C., Weinberg, D. H., Hernquist, L. \& Katz, N.,  1997a,
in  Proceedings of the 18th Texas Symposium on
Relativistic Astrophysics, eds. A. Olinto, J. Frieman, \& D. Schramm,
(Singapore: World Scientific), (astro-ph/9701166)

\bibitem[Croft et al.\ 1997b]{cwkh97b}
Croft, R.A.C., Weinberg, D.H., Katz, N., Hernquist, L., 1997b, \apj, in press
(astro-ph/9611053)

\bibitem[Croft et al.\ 1997c]{cwkh97c}
Croft, R. A. C., Weinberg, D. H., Hernquist, L., \& Katz, N., 1997c,
\apj, submitted (astro-ph/9708018)

\bibitem[Crotts \& Fang 1997]{crotts97}
Crotts, A. P. S., \& Fang, Y. 1997, \apj, submitted (astro-ph/9702185)

\bibitem[Dav\'e et al.\ 1997]{dhwk97}
Dav\'e, R., Hernquist, L., Weinberg, D. H., \& Katz, N. 1997, \apj, 477, 21

\bibitem[Davis et al.\ 1985]{davis85}
Davis, M., Efstathiou, G., Frenk, C. S., \& White,
S. D. M. 1985, \apj, 292, 371

\bibitem[Dinshaw et al.\ 1994]{dinshaw94}
Dinshaw, N., Impey, C. D., Foltz, C. B., Weymann, R. J., \&
Chaffee, F. H. 1994, \apj, 437, L87

\bibitem[Dinshaw et al.\ 1995]{dinshaw95}
Dinshaw, N., Foltz, C. B., Impey, C. D., Weymann, R. J., \&
Morris, S. L. 1995, Nature, 373, 223

\bibitem[Doroshkevich 1970]{doroshkevich70}
Doroshkevich, A. G. 1970, Astrofisika, 6, 581 [Engl. trans.
in 1973, Astrophysics, 6, 320]

\bibitem[Efstathiou \& Eastwood 1981]{efstathiou81}
Efstathiou, G., \& Eastwood, J. W. 1981, \mnras, 194, 503

\bibitem[Efstathiou et al.\ 1988]{efstathiou88}
Efstathiou, G., Frenk, C. S., White, S. D. M., \& Davis, M. 1988,
\mnras, 235, 715

\bibitem[Evrard, Summers, \& Davis 1994]{evrard94}
Evrard, A.E., Summers, F.J., \& Davis, M. 1994, \apj, 422, 11

\bibitem[{Fisher et al.\ }1994]{fisher94}
Fisher, K.B., Davis, M., Strauss, M.A., Yahil, A., \& Huchra, J.P.
1994, \mnras, 267, 927

\bibitem[Fry \& Melott 1985]{fry85}
Fry, J. N., \& Melott, A. L. 1985, \apj, 292, 395

\bibitem[Gardner et al.\ 1997]{gkhw97}
Gardner, J. P., Katz, N., Hernquist, L., \& Weinberg, D. H. 1997,
\apj, 484, 31

\bibitem[Gelb \& Bertschinger 1994]{gelb94}
Gelb, J. M., \& Bertschinger, E. 1994, \apj, 436, 491

\bibitem[Gingold \& Monaghan 1977]{gingold77}
Gingold, R.A., \& Monaghan, J.J. 1977, \mnras, 181, 375

\bibitem[Gnedin 1997]{gnedin97}
Gnedin, N. Y. 1997, \mnras, submitted (astro-ph/9706286)

\bibitem[Gnedin \& Hui 1997]{gh97}
Gnedin, N. Y. \& Hui, L. 1997, \mnras, submitted (astro-ph/9706219)

\bibitem[Gnedin \& Ostriker 1997]{go97}
Gnedin, N. Y., \& Ostriker, J. P. 1997, \apj, 486, 581

\bibitem[Gooding et al.\ 1992]{gooding92}
Gooding, A. K., Park, C., Spergel, D. N., Turok, N.,
Gott, J. R. 1992, \apj, 393, 42

\bibitem[Gott, Turner, \& Aarseth 1979]{gott79}
Gott, J. R., Turner, E. L., \& Aarseth, S. J. 1979, \apj, 234, 13

\bibitem[Gott, Weinberg, \& Melott 1987]{gwm87}
Gott, J. R., Weinberg, D. H., \& Melott, A. L. 1987, \apj, 319, 1

\bibitem[{Gunn \& Peterson }1965]{gunn65}
Gunn, J.E., \& Peterson, B.A. 1965, \apj, 142, 1633

\bibitem[Haardt \& Madau 1996]{haardt96}
Haardt, F., \& Madau, P. 1996, \apj, 461, 20

\bibitem[Haehnelt, Steinmetz, \& Rauch 1996]{haehnelt96}
Haehnelt, M. G., Steinmetz, M., \& Rauch, M. 1996, \apj, 465, L95

\bibitem[Haehnelt, Steinmetz, \& Rauch 1997]{haehnelt97}
Haehnelt, M. G., Steinmetz, M., \& Rauch, M. 1997, \apj, submitted
(astro-ph/9706201)

\bibitem[Haiman \& Loeb 1997]{haiman97}
Haiman, Z., \& Loeb, A. 1997, \apj, 483, 21

\bibitem[Hamilton, Gott, \& Weinberg 1986]{hgw86}
Hamilton, A. J. S., Gott, J. R., \& Weinberg, D. H. 1986,
\apj, 309, 1

\bibitem[Hellsten et al.\ 1997]{hdhkw97}
Hellsten, U., Dav\'e, R., Hernquist, L., Katz, N., \& Weinberg, D. H. 1997a,
\apj, submitted (astro-ph/9708090)

\bibitem[Hellsten et al.\ 1997]{hdhwk97}
Hellsten, U., Dav\'e, R., Hernquist, L., Weinberg, D. H., \& Katz, N. 1997b,
\apj, in press (astro-ph/9701043)

\bibitem[Hernquist 1987]{hernquist87}
Hernquist, L. 1987, ApJS, 64, 715

\bibitem[Hernquist \& Katz 1989]{hernquist89}
Hernquist, L., \& Katz, N. 1989, \apjs, 70, 419

\bibitem[Hernquist et al.\ 1996]{hkwm96}
Hernquist L., Katz, N., Weinberg, D.H., \&
Miralda-Escud\'e, J. 1996, \apj, 457, L5

\bibitem[Hui \& Gnedin 1997]{hg97}
Hui, L., \& Gnedin, N. 1997, \mnras, submitted (astro-ph/9612232)

\bibitem[Hui, Gnedin, \& Zhang 1997]{hui97}
Hui, L., Gnedin, N., \& Zhang, Y. 1997, \apj, in press (astro-ph/9608157)

\bibitem[Katz 1992]{katz92}
Katz, N. 1992, \apj, 391, 502

\bibitem[Katz \& Gunn 1991]{katz91}
Katz, N., \& Gunn, J. E. 1991, \apj, 377, 365

\bibitem[Katz et al.\ 1992]{khw92}
Katz, N., Hernquist, L., \& Weinberg, D. H. 1992, \apj, 399, L109

\bibitem[Katz, Weinberg, \& Hernquist 1996a]{kwh96}
Katz, N., Weinberg D.H., \& Hernquist, L. 1996a, \apjs, 105, 19 (KWH)

\bibitem[Katz et al.\ 1996b]{kwhm96}
Katz, N., Weinberg D.H., Hernquist, L., \& Miralda-Escud\'e, J. 1996b,
\apj, 457, L57

\bibitem[Kauffmann, White, \& Guideroni 1993]{kauffmann93}
Kauffmann, G., White, S. D. M., \& Guideroni, B. 1993, \mnras, 264, 201

\bibitem[Kim \& Strauss 1997]{kim97a}
Kim, R. S., \& Strauss, M. A. 1997, \apj, submitted (astro-ph/9702144)

\bibitem[Kim et al.\ 1997]{kim97}
Kim, T. S., Hu, E. M., Cowie, L. L., \& Songaila, A. 1997, \aj, 114, 1

\bibitem[Klypin, Gottl\"ober, \& Kravtsov 1997]{klypin97}
Klypin, A., Gottl\"ober, S., \& Kravtsov, A. V. 1997, \apj, submitted
(astro-ph/9708191)

\bibitem[Klypin \& Shandarin 1983]{klypin83}
Klypin, A. A., \& Shandarin, S. F. 1983, \mnras, 204, 891

\bibitem[Lacey \& Cole 1994]{lacey94}
Lacey, C., \& Cole, S. 1994, \mnras, 271, 676

\bibitem[Little, Weinberg, \& Park 1991]{lwp91}
Little, B., Weinberg, D. H., \& Park, C. 1991, \mnras, 253, 295

\bibitem[Lowenthal et al.\ 1997]{lowenthal97}
Lowenthal, J. D., Koo, D. C., Guzman, R., Gallego, J., Phillips, A. C.,
Faber, S. M., Vogt, N. P., Illingworth, G. D., \& Gronwall, C. 1997,
\apj, 481, 673

\bibitem[Lucy 1977]{lucy77}
Lucy, L. 1977, \aj, 82, 1013

\bibitem[{Lynds }1971]{lynds71}
Lynds, C.R. 1971, \apj , 164, L73

\bibitem[Madau 1997]{madau97}
Madau, P. 1997, in The Hubble Deep Field, eds. M. Livio,
S. M. Fall, \& P. Madau (Cambridge: Cambridge University Press), in press

\bibitem[McGill 1990]{mcgill90}
McGill, C. 1990, \mnras, 242, 544

\bibitem[Melott 1987]{melott87}
Melott, A. L. 1987, \mnras, 228, 1001

\bibitem[Melott \& Scherrer 1987]{melott87a}
Melott, A. L., \& Scherrer, R. J. 1987, Nature, 328, 691

\bibitem[Melott \& Shandarin 1993]{melott93}
Melott, A. L., \& Shandarin, S. F. 1993, \apj, 410, 469

\bibitem[Meyer \& York 1987]{meyer87}
Meyer, D. M., \& York, D.G. 1987, \apj, 315, L5

\bibitem[Miralda-Escud\'e et al.\ 1996]{miralda96}
Miralda-Escud\'e J., Cen R., Ostriker, J.P., \& Rauch, M. 1996, \apj, 471, 582

\bibitem[Miralda-Escud\'e \& Rees 1997]{miralda97}
Miralda-Escud\'e J., \& Rees, M. J. 1997, \apj, 478, L57

\bibitem[Moore, Katz, \& Lake 1996]{moore96}
Moore, B., Katz, N., \& Lake, G. 1996, \apj, 457, 455

\bibitem[M\"ucket et al.\ 1996]{mucket96}
M\"ucket, J. P., Petitjean, P., Kates, R. E., \& Riediger, R. 1996,
A\&A, 308, 17

\bibitem[Navarro, Frenk, \& White 1995]{navarro95}
Navarro, J. F., Frenk, C. S., \& White, S. D. M. 1995, \mnras, 275, 56

\bibitem[Navarro, Frenk, \& White 1996]{navarro96}
Navarro, J., Frenk, C. S., \& White, S. D. M. 1996, \apj, 462, 563

\bibitem[Navarro \& Steinmetz 1997]{navarro97}
Navarro, J. F., \& Steinmetz, M. 1997, \apj, 478, 13

\bibitem[Park 1990]{park90}
Park, C. 1990, \mnras, 242, L59

\bibitem[Park \& Gott 1991]{park91}
Park, C., \& Gott, J. R. 1991, \mnras, 249, 288

\bibitem[Petitjean, M\"ucket, \& Kates 1995]{petitjean95}
Petitjean, P., M\"ucket, J. P., \& Kates, R. E. 1995, A\&A, 295, L9

\bibitem[Press \& Rybicki 1993]{pr93}
Press, W. H., \& Rybicki, G. B. 1993, \apj, 414, 64

\bibitem[Protogeros \& Scherrer 1997]{protogeros97}
Protogeros, Z. A. M., \& Scherrer, R. J. 1997, \mnras, 284, 425

\bibitem[Rauch, Haehnelt, \& Steinmetz 1997]{rhs97}
Rauch, M., Haehnelt, M. G., \& Steinmetz, M. 1997a, \apj, 481, 601

\bibitem[Rauch et al.\ 1997b]{rauch97}
Rauch, M., Miralda-Escud\'e, J., Sargent, W. L. W., Barlow, T. A.,
Weinberg, D. H., Hernquist, L., Katz, N., Cen, R., \& Ostriker, J. P.,
1997b, \apj, in press (astro-ph/9612245)

\bibitem[Rugers \& Hogan 1996]{rugers96}
Rugers, M., \& Hogan, C.J. 1996a, \aj, 111, 2135

\bibitem[Rugers \& Hogan 1996]{rugers96b}
Rugers, M., \& Hogan, C.J., 1996b, \apj, 459, L1

\bibitem[{Sargent et al.\ }1980]{sargent80}
Sargent, W.L.W., Young, P.J., Boksenberg, A. \& Tytler, D.  1980, \apjs, 42, 41

\bibitem[Saslaw 1989]{saslaw89}
Saslaw, W. C. 1989, \apj, 341, 548

\bibitem[Scherrer, Melott, \& Bertschinger 1989]{scherrer89}
Scherrer, R. J., Melott, A. L., \& Bertschinger, E. 1989, Phys Rev Lett, 62, 379

\bibitem[Songaila \& Cowie 1996]{songaila96}
Songaila, A. \& Cowie, L.L. 1996, \aj, 112, 335

\bibitem[Steidel et al.\ 1996]{steidel96}
Steidel, C. C., Giavalisco, M., Pettini, M., Dickinson, M.,
\& Adelberger, K. L. 1996, \apj, 462, L17

\bibitem[Steinmetz \& M\"uller 1994]{steinmetz94}
Steinmetz, M., \& M\"uller, E. 1994, A\&A, 281, L97

\bibitem[Steinmetz \& M\"uller 1995]{steinmetz95}
Steinmetz, M., \& M\"uller, E. 1995, \mnras, 276, 549

\bibitem[Tytler et al.\ 1996]{tytler96}
Tytler, D., Fan, X.M, \& Burles, S. 1996, Nature, 381, 207

\bibitem[Vedel, Hellsten, \& Sommer-Larsen 1994]{vedel94}
Vedel, H., Hellsten, U., \& Sommer-Larsen, J. 1994, \mnras, 271, 743

\bibitem[Weinberg \& Cole 1992]{wc92}
Weinberg, D. H., \& Cole, S. 1992, \mnras, 259, 652

\bibitem[Weinberg et al.\ 1997]{wmhk97}
Weinberg, D.H., Miralda-Escud\'{e}, J., Hernquist, L., \& Katz, N., 1997,
\apj, in press (astro-ph 9701012)

\bibitem[{White }1996]{white96}
White, S. D. M. 1996, in Cosmology and Large Scale Structure, 
eds. R. Schaefer, J. Silk, M. Spiro, \& 
J. Zinn-Justin, (Dordrecht: Elsevier), (astro-ph/9410043)

\bibitem[White, Frenk, \& Davis 1983]{white83}
White, S. D. M., Frenk, C. S., \& Davis, M. 1983, \apj, 274, L1

\bibitem[White et al.\ 1987]{white87}
White, S. D. M., Frenk, C. S., Davis, M., \& Efstathiou, G. 1987, \apj, 313, 505

\bibitem[White \& Rees 1978]{white78}
White, S. D. M., \& Rees, M. J. 1978, \mnras, 183, 341

\bibitem[Womble, Sargent, \& Lyons 1996]{womble96}
Womble, D.S., Sargent, W.L.W., \& Lyons, R.S. 1996, in
Cold Gas at High Redshift, eds. M. Bremer et al., (Dordrecht: Kluwer),
(astro-ph/9511035)

\bibitem[Zel'dovich 1970]{zeldovich70}
Zel'dovich, Y. B. 1970, A\&A, 5, 84

\bibitem[Zhang, Anninos, \& Norman 1995]{zhang95}
Zhang, Y., Anninos, P., \& Norman, M.L. 1995, \apj, 453, L57

\bibitem[Zhang et al.\ 1997]{zhang97}
Zhang, Y., Anninos, P., Norman, M. L., \& Meiksin, A. 1997, \apj, 485, 496

\bibitem[Zuo \& Bond 1994]{zuo94}
Zuo, L., \& Bond, J. R. 1994, \apj, 423, 73
}
\end{thebibliography}
\end{document}